\documentclass[10pt,aps,pre,twocolumn,nofootinbib,floatfix]{revtex4-2}

\usepackage{amssymb,amsmath}
\usepackage{graphicx}
\usepackage[usenames,dvipsnames,svgnames,table]{xcolor}
\usepackage{hyperref}
\usepackage{siunitx}
\usepackage{bm}
\usepackage{textcomp}
\usepackage{gensymb}
\usepackage{placeins}
\usepackage{physics}
\usepackage{nicefrac}
\usepackage[caption=false]{subfig}
\hypersetup{
	pdfauthor={},     		 
	colorlinks=true,       	 
	linkcolor=blue,          
	citecolor=red           
}

\newcommand{\kT}{k_{\mathrm{B}}T}
\newcommand{\X}{X}
\newcommand{\Y}{Y}
\newcommand{\x}{$x$}
\newcommand{\y}{$y$}
\newcommand{\ex}{x} 
\newcommand{\ey}{y}
\newcommand{\tc}{transduced capacity}

\newcommand{\dX}{\Delta \rm X}
\newcommand{\bX}{\bar{X}}
\newcommand{\dx}{$\Delta x$}
\newcommand{\bx}{$\bar{x}$}
\newcommand{\dex}{\Delta x}
\newcommand{\bex}{\bar{x}}

\newcommand{\add}[1]{#1} 
\usepackage[normalem]{ulem}
\usepackage{cancel}
\newcommand{\stkout}[1]{}

\begin{document}
\title{Internal energy and information flows mediate input and output power in bipartite molecular machines}
\author{Emma Lathouwers}
\altaffiliation{\add{Current address: AbCellera Biologics Inc., Vancouver, BC, Canada, V5Y0A1}}
\author{David A.\ Sivak}
\email{dsivak@sfu.ca}
\affiliation{Department of Physics, Simon Fraser University, Burnaby, BC, Canada, V5A1S6}
\date{\today}

\begin{abstract}
Microscopic biological systems operate far from equilibrium, are subject to strong fluctuations, and are composed of many coupled components with interactions varying in nature and strength. Researchers are actively investigating the general design principles governing how biomolecular machines achieve effective free-energy transduction in light of these challenges. We use a model of two strongly coupled stochastic rotary motors to explore the effect of coupling strength between components of a molecular machine. We observe prominent thermodynamic characteristics at intermediate coupling strength, near that which maximizes output power: a maximum in power and information transduced from the upstream to the downstream system, and equal subsystem entropy production rates. These observations are unified through a bound on the machine's input and output power, which accounts for both the energy and information transduced between subsystems.
\end{abstract}

\maketitle

\section{Introduction} \label{sec:intro}

    Living things must stay out of thermal equilibrium in order to persist~\cite{Schrodinger1944}. 
    Molecular machines play an important role in maintaining the far-from-equilibrium conditions and make use of the same conditions to perform various tasks~\cite{Kolomeisky2007,Brown2017,Brown2019}.
    Molecular machines transduce between different nonequilibrium stores of free energy such as out-of-equilibrium concentrations of chemical reactants and products, spatial concentration gradients, and elastic mechanical energy. 
    They perform these functions in a crowded cellular environment subject to constant stochastic fluctuations, and they typically consist of many coupled components with interactions that vary in nature and strength.
    How they effectively transduce energy among their relatively flexibly linked degrees of freedom when they experience strong fluctuations is not well understood.
    
    F$_{\rm O}$F$_1$-ATP synthase~\cite{Nelson2004} is an experimentally heavily studied motor that synthesizes adenosine triphosphate (ATP), the most widespread energy currency within the cell. The F$_{\rm O}$ part of the motor harnesses a proton gradient across a membrane to rotate a central crankshaft, inducing a conformational change in the F$_1$ part, which then catalyzes the synthesis of ATP~\cite{Boyer1997, Yoshida2001, Junge2015}. As a multipart motor with different parts driven by different chemical reservoirs, linked by a particularly simple mechanical coupling, capable of achieving high efficiency~\cite{Silverstein2014} and high speed~\cite{Ueno2005} despite slip~\cite{Toyabe2011,Feniouk1999} due to somewhat flexible coupling~\cite{Sielaff2008,Okuno2010}, ATP synthase forms a model system for biomolecular energy and information transmission.

    Inspired by the example of ATP synthase and its particularly simple coarse-grained architecture, we use a model 
    \add{(originally formulated in \cite{Lathouwers2020})}
    of two coupled subsystems to study the connections between functional output and the internal energy and information flows between different components of a strongly fluctuating microscopic system.
    Previously, we found that intermediate-strength coupling maximizes output power (in many contexts a primary function of the system) while the efficiency is maximized at tight coupling~\cite{Lathouwers2020}.
    In this paper, we reveal apparently coinciding phenomena at intermediate-strength coupling of maximal output power, maximal \tc~(energy plus information flow), and equal subsystem entropy production rates.
    The coinciding features are unified by local second laws revealing that the \tc~lower bounds the 
    \stkout{output}
    \add{input}
    power and upper bounds the 
    \stkout{input}
    \add{output}
    power, with subsystem entropy production rates quantifying the dissipations and hence the capacity losses in each subsystem.

\section{Model} \label{sec:model}

\begin{figure}[htb]
        \centering 
        \includegraphics[width=\columnwidth]{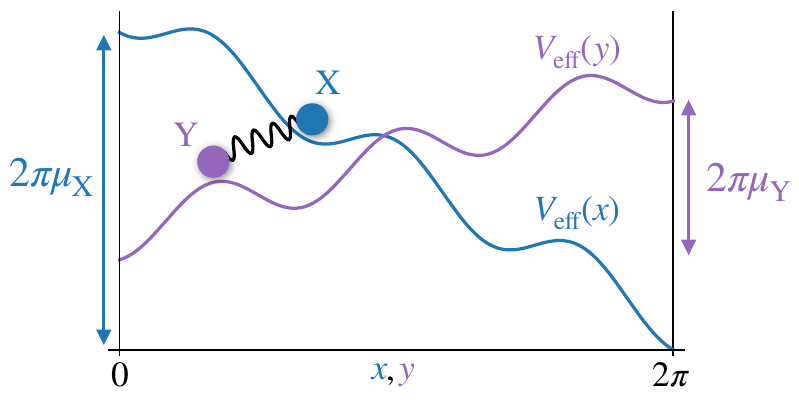}
        \caption{Effective energy landscapes experienced by each subsystem.  
        Each subsystem diffuses on a periodic 
        effective 
        energy landscape and is coupled (depicted by a connecting spring) to the other subsystem. 
        The tilt of each landscape reflects the effect of its respective chemical driving force.
        Figure adapted from \cite{Lathouwers2020}.}
        \label{fig:sketch}
    \end{figure}

    \stkout{The two subsystems} 
    \add{Two machine components}
    \X~and \Y~\add{(e.g., representing F$_{\rm O}$ and F$_1$)} 
    are modeled as energetically coupled subsystems, each performing a biased random walk on a periodic energy landscape 
    \add{(representing a single mechanochemical cycle)}
    driven by a constant chemical driving force
    \add{due to the chemical species (e.g., protons or ATP and its hydrolysis products) that are externally maintained at constant out-of-equilibrium concentrations by other cellular machinery}. 
    A sketch of the system is shown in Fig.~\ref{fig:sketch}.
    The two-dimensional energy landscape of the joint system has a contribution for each individual subsystem and a 
    \stkout{coupling} 
    contribution
    \add{due to mechanical coupling (e.g., ATP synthase's crankshaft) between subsystems}:
    \begin{subequations} \label{eq:energylandscape} 
        \begin{gather}
            \hspace{-0.3em} V(\ex, \ey) = V_{\rm \X}(\ex) + V_{\rm \Y}(\ey) + V_{\rm couple}(\ex, \ey) \\
            \begin{aligned}
                \hspace{2em} = &-\tfrac{1}{2} E^\ddagger \cos{3 \ex} - \tfrac{1}{2} E^\ddagger \cos{3 \ey} \\
                &- \tfrac{1}{2}E_{\rm couple} \cos{ \left( \ex - \ey \right) } .
            \end{aligned}
        \end{gather}
    \end{subequations}
    Here \x~and \y~are the respective states of the subsystems, each with periodic boundary conditions ($\ex, \ey \in [0, 2 \pi)$); $E^\ddagger$ is the barrier height of each (untilted) subsystem-specific landscape; and $E_{\rm couple}$ is the strength of the coupling between subsystems. 
    The cosine potentials with 3 metastable states incorporate straightforward periodic boundary conditions.
    
    The upstream system \X~is driven by a constant chemical driving force $\mu_{\rm \X}$, and the downstream system \Y~is pushed by constant chemical driving force $\mu_{\rm \Y}$.
    Chemical driving forces are constrained to be in opposition ($\mu_{\rm \X} > 0$, $\mu_{\rm \Y} < 0$), so \X~is pushed clockwise and \Y~is pushed counter-clockwise.
    Without loss of generality, the `upstream' \X~subsystem is more strongly driven ($|\mu_{\rm \X}| > |\mu_{\rm \Y}|$).
    The two subsystems thus together form an isothermal work-to-work converter, transducing between different stores of free energy each connected to a single subsystem.
    
    In the absence of coupling ($\beta E_{\rm couple}=0$), \X~is only subject to its subsystem-specific effective landscape $V_{\rm eff}(\ex) 
    \add{\equiv V_{\rm \X}(x) - \mu_{\rm \X} \ex}$,
    the energy landscape $V_{\rm \X}(x)$ `tilted' by the chemical driving force $\mu_{\rm \X}$ applied to it (and similarly for \Y). 
    The coupling between \X~and \Y~favors the subsystems to be close together, thereby favoring movement in the same direction.
    Each subsystem can on average move in either direction, with the sign depending on the strength of driving forces and coupling.  
    If the subsystems move in the same direction (on average) while being pushed in opposite directions by the applied driving forces, the system as a whole transduces energy from one work reservoir to the other.
    
    The joint probability distribution $p(\ex,\ey;t)$ describing the overdamped, stochastic system evolves according to the Smoluchowski equation~\cite[Ch.~VIII]{VanKampen1992},
    \begin{align} \label{eq:2dFPE}
    	&\pdv{}{t} p(\ex,\ey;t) = -\pdv{\mathcal{J}_{\rm \X}(\ex, \ey)}{\ex} - \pdv{\mathcal{J}_{\rm \Y}(\ex, \ey)}{\ey} \ ,
    \end{align}
    with probability currents
    \begin{subequations} \label{eq:flux}
        \begin{align}
        	\mathcal{J}_{\rm \X}(\ex, \ey) &= \dfrac{1}{\zeta} \left[
        	\left(\pdv{V}{\ex} - \mu_{\rm \X} \right) + \frac{1}{\beta} \pdv{}{\ex} \right] p(\ex, \ey; t) \\
        	\mathcal{J}_{\rm \Y}(\ex, \ey) &= \dfrac{1}{\zeta} \left[
        	\left(\pdv{V}{\ey} - \mu_{\rm \Y} \right) + \frac{1}{\beta} \pdv{}{\ey} \right] p(\ex, \ey; t) \ ,
        \end{align}
    \end{subequations}
    friction coefficient $\zeta$, and $\beta \equiv 1/(\kT)$ for temperature $T$ and Boltzmann's constant $k_{\rm B}$. 
    The diagonal diffusion matrix embodied in the second terms within the square brackets means that each subsystem is subject to independent noise and thus the dynamics are bipartite~\cite{Chetrite2019_Information}, simplifying the identification of thermodynamic flows. 
    
    The system is initialized with the equilibrium Boltzmann distribution for no chemical driving forces ($\mu_{\rm \X}=0=\mu_{\rm \Y}$),
    \begin{align}
        p(x, y; t=0) \propto \exp \left[ -\beta V(x, y) \right] \ .
    \end{align}
    The probability distribution is dynamically evolved using \add{a custom forward-time centered-space~\cite{Press2007} finite-difference code~\cite{Lucero2020} to approximate} Eq.~\eqref{eq:2dFPE} until it reaches steady state.
    \add{(Appendix~\ref{app:compDetails} provides computational details.)}
    \stkout{as quantified by the total variation distance not changing significantly,}
    The resulting steady-state probability distribution determines the steady-state probability currents~\eqref{eq:flux}, which in turn dictate the system's energy flows.
    
    The subsystems are \emph{strongly coupled} in the sense that there is a (significant) energy associated with the interaction between the subsystems; this is in contrast with thermodynamically large systems where the energy at the interface between the system and environment can generally be safely neglected.
    The subsystems are \emph{tightly coupled} when the coupling potential is so strong that they move in lockstep and can be described by a single coordinate (in this model, 
    \add{only the case} 
    when $\beta E_{\rm couple} \gg 1$).

\section{Thermodynamics} \label{sec:theory}

    \begin{figure}[th]
        \centering
        \includegraphics[width=\columnwidth]{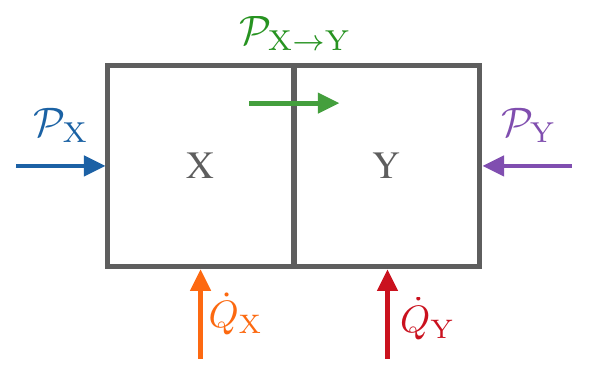}
        \caption{Energy flows in and out of the subsystems. Arrows indicate the directions of positive flow.}
        \label{fig:directionality}
    \end{figure}
    
    \stkout{We} 
    \add{Unlike tight coupling between subsystems which we do not generally obtain in our model, we do (by construction)}
    assume tight coupling between the chemical reservoirs and mechanical motion of the respective subsystems, i.e., that each mechanical movement $d\ex$ ($d\ey$) of subunit X (Y) is accompanied by transfer of energy $\mu_{\rm \X}\,d\ex$ ($\mu_{\rm \Y}\,d\ey$) from the respective chemical reservoir. This effectively consolidates all interfaces of finite coupling strength across the whole system into the single coupling between \X~and \Y.
    Thus the input power $\mathcal{P}_{\rm \X}$ and output power $\mathcal{P}_{\rm \Y}$ are each the integrated probability current multiplied by its respective chemical driving force,
    \begin{subequations}
        \begin{align}
    		\mathcal{P}_{\rm \X} &= \mu_{\rm \X} \int d\ex \int d\ey \, \mathcal{J}_{\rm \X}(\ex, \ey) \\
    		\mathcal{P}_{\rm \Y} &= \mu_{\rm \Y} \int d\ex \int d\ey \, \mathcal{J}_{\rm \Y}(\ex, \ey) \ .
	    \end{align}
    \end{subequations}
	Figure~2 illustrates all internal and external energy flows.
	
    The bipartite dynamics encapsulated in \eqref{eq:flux} identifies the heat flow $\dot{Q}$ associated with the dynamics of a particular subsystem as the change in the system potential energy due to the dynamics of that particular subsystem, less the associated power: 
	\begin{subequations}
	    \begin{equation}
    	    \begin{aligned}
        		\dot{Q}_{\rm \X} = &\int d\ex \int d\ey \, \mathcal{J}_{\rm \X}(\ex, \ey) \, \partial_{\ex} V(\ex, \ey) \\
        		&- \mu_{\rm \X} \int d\ex \int d\ey \, \mathcal{J}_{\rm \X}(\ex, \ey) 
        	\end{aligned}
        \end{equation}
        \begin{equation}
    	    \begin{aligned}
        		\dot{Q}_{\rm \Y} = &\int d\ex \int d\ey \, \mathcal{J}_{\rm \Y}(\ex, \ey) \, \partial_{\ey} V(\ex, \ey) \\
        		&- \mu_{\rm \Y} \int d\ex \int d\ey \, \mathcal{J}_{\rm \Y}(\ex, \ey) \ .
        	\end{aligned}
    	\end{equation}
	\end{subequations}
	
	The transduced powers (rates of control work in \cite{Barato2017}) are the respective changes in energy due to the dynamics of each subsystem, 
	\begin{subequations} \label{eq:transpower}
	    \begin{align}
    		\mathcal{P}_{\rm \X \to \Y} &= \int d\ex \int d\ey \, \mathcal{J}_{\rm \X}(\ex, \ey) \, \partial_{\ex} V(\ex, \ey) \ , \\
    		\mathcal{P}_{\rm \Y \to \X} &= \int d\ex \int d\ey \, \mathcal{J}_{\rm \Y}(\ex, \ey) \, \partial_{\ey} V(\ex, \ey) \ .
    	\end{align}
	\end{subequations}
	
	At steady state (where average system energy change vanishes, $\tfrac{d}{dt} \int dx \, dy \, p(x,y;t) V(x,y) = 0$), the average rates of work (average powers) done by one subsystem on the other are equal magnitude and opposite sign, $\mathcal{P}_{\rm \X \to \Y} = - \mathcal{P}_{\rm \Y \to \X}$, giving ensemble-average first laws for each subsystem:
    \begin{subequations} \label{eq:FLo} 
        \begin{align}
            \mathcal{P}_{\rm \X} + \dot{Q}_{\rm \X} - \mathcal{P}_{\rm \X \to \Y} &= 0 \\
    		\mathcal{P}_{\rm \Y} + \dot{Q}_{\rm \Y} + \mathcal{P}_{\rm \X \to \Y} &= 0 \ . \label{eq:FL1}  
        \end{align}
	\end{subequations}
	
	The subsystem-specific entropy production rates (EPRs) in a bipartite system are~\cite{Horowitz2015} 
	\begin{subequations}
        \begin{align} 
            \dot{\Sigma}_{\rm \X} &= d_t H(\X) + \dot{\Sigma}_{\rm \X}^{\rm res} - \dot{I}_{\rm \X} \label{eq:epr1} \\
            \dot{\Sigma}_{\rm \Y} &= d_t H(\Y) + \dot{\Sigma}_{\rm \Y}^{\rm res} - \dot{I}_{\rm \Y} \label{eq:epr2} \ ,
        \end{align}
    \end{subequations}
    for marginal 
    \stkout{entropies $H(\X)$ and $H(\Y)$. They} 
    \add{Shannon entropy $H(\X) \equiv -\int d\x \, p(\x) \log p(\x)$ and similarly for \Y~\cite{Cover:2006:Book}. Eqs.~\eqref{eq:epr1} and \eqref{eq:epr2}}
    sum to the total system entropy production rate. 
    In our system, the change in entropy due to contact with environmental reservoirs is due to contact with a thermal reservoir, hence $\dot{\Sigma}_{\rm \X}^{\rm res} = -\beta \dot{Q}_{\rm \X}$ and $\dot{\Sigma}_{\rm \Y}^{\rm res} = -\beta \dot{Q}_{\rm \Y}$.
    
    The last component of the EPR is the information flow~\cite{Allahverdyan2009} (also known as the learning rate~\cite{Barato2014b,Brittain2017}), quantifying the change in mutual information between the subsystems due to changes in one subsystem,
    \begin{subequations} \label{eq:infoflow}
        \begin{align}
            \dot{I}_{\rm \X} &= \partial_{\tau} I[X(t+\tau);Y(t)] |_{\tau \to 0} \\
            \dot{I}_{\rm \Y} &= \partial_{\tau} I[X(t);Y(t+\tau)] |_{\tau \to 0} \ .
        \end{align}
    \end{subequations}
    The mutual information $I[\X,\Y] \equiv H(\X)+H(\Y)-H(\X,\Y) = H(\Y)-H(\Y|\X) = H(\X)-H(\X|\Y)$ between random variables $X$ and $Y$ (for joint entropy $H(\X,\Y)$ and conditional entropies $H(\X|\Y)$ and $H(\Y|\X)$) quantifies the expected reduction in the uncertainty of one variable upon learning the other variable~\cite{Cover:2006:Book}. It can be written in terms of the marginal distributions $p(\ex)$ and $p(\ey)$ and joint distribution $p(\ex, \ey)$ as
    \begin{align} \label{eq:MI}
        I[\text{\X}; \text{\Y}] = \int d\ex \int d\ey \, p(\ex,\ey) \, \log \frac{p(\ex, \ey)}{p(\ex) p(\ey)} \ .
    \end{align}
    The information-flow definition~\eqref{eq:infoflow} can be manipulated~\cite{Horowitz2015} to a form that highlights the similarity to the transduced power~\eqref{eq:transpower}:
    \begin{subequations} \label{eq:infoflow2}
        \begin{align} 
            \dot{I}_{\rm \X} &= \int d\ex \int d\ey \, \mathcal{J}_{\rm \X}(\ex, \ey) \partial_{\ex} \log p(\ey|\ex)   \\
            \dot{I}_{\rm \Y} &= \int d\ex \int d\ey \, \mathcal{J}_{\rm \Y}(\ex, \ey) \partial_{\ey} \log p(\ex|\ey)   \ .
        \end{align} 
    \end{subequations}
    
    At steady state, each subsystem's Shannon entropy does not change ($d_t H(\X) = 0 = d_t H(\Y)$) and the joint system entropy does not change ($d_t H(\X,\Y) = 0$), so $\dot{I}_{\rm \X} = - \dot{I}_{\rm \Y}$.   
    Each subsystem separately obeys the second law (each subsystem entropy production is nonnegative), hence at steady state
    \begin{subequations} \label{eq:EPR}
        \begin{align} \label{eq:EPRo}
    		0 \leq \dot{\Sigma}_{\rm \X} &= -\beta \dot{Q}_{\rm \X} - \dot{I}_{\rm \X} \\
    		0 \leq \dot{\Sigma}_{\rm \Y} &= -\beta \dot{Q}_{\rm \Y} + \dot{I}_{\rm \X} \label{eq:EPR1} \ .
    	\end{align}
    \end{subequations}

\section{Results} \label{sec:results}
    
    \subsection{Energy flows}
\begin{figure}[!th]
            \centering
            \includegraphics[width=\columnwidth]{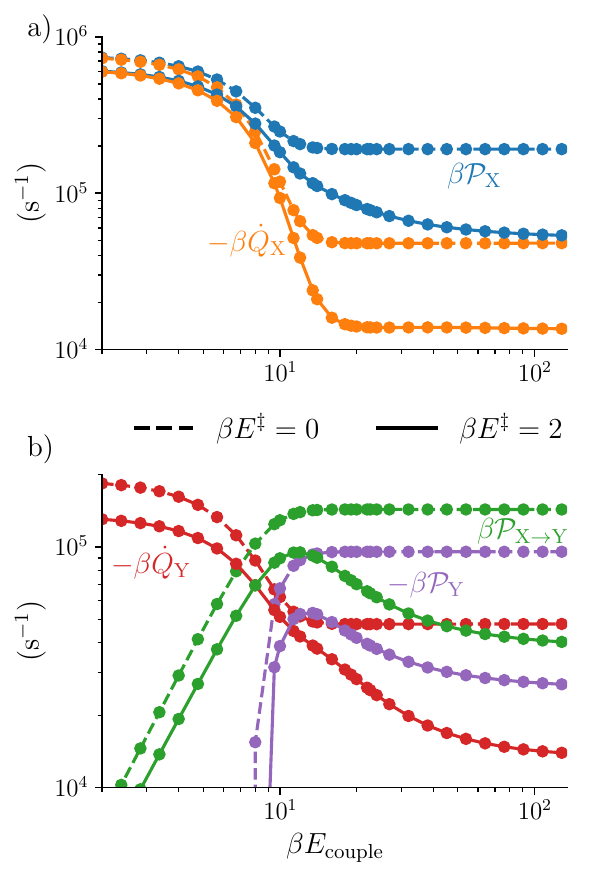}
            \caption{a) Input power $\beta\mathcal{P}_{\rm \X}$ and heat flow $\beta\dot{Q}_{\rm \X}$ due to the dynamics of \X, and b) output power $\beta\mathcal{P}_{\rm \Y}$, heat flow $\beta\dot{Q}_{\rm \Y}$ due to the dynamics of \Y, and transduced power $\beta\mathcal{P}_{\rm \X \to \Y}$, all as functions of coupling strength. Dashed curves: barrierless system ($\beta E^\ddagger=0$). Solid curves: system with $n = 3$ barriers of height $\beta E^\ddagger = 2$. Chemical driving forces are $\beta \mu_{\rm \X} = 4\, \rm rad^{-1}$ and $\beta \mu_{\rm \Y} = -2\, \rm rad^{-1}$.
            \label{fig:energyflow}
            }
        \end{figure}

        Figure~\ref{fig:energyflow} shows steady-state energy flows through the system as a function of coupling strength $E_{\rm couple}$ for a barrierless system and a system with energy barriers.
        (Systems with barriers hold more intrinsic interest, but barrierless systems help isolate the contribution of subunit coupling.)
        At low coupling strength, the subsystem's dynamics is dominated by its intrinsic energy landscape and associated driving force: each subsystem on average moves in the direction of its respective driving force, so energy is not transduced from one subsystem to the other, but rather is all dissipated as heat to the environment, indicated by the relatively high values of $-\dot{Q}_{\rm \X}$ and $-\dot{Q}_{\rm \Y}$.
        As the coupling strength increases, energy is increasingly transduced from the stronger-driven upstream subsystem \X~to the weaker-driven downstream subsystem \Y~($\mathcal{P}_{\rm \X \to \Y}$ increases), decreasing the dissipated heat from each subsystem and changing the sign of the output power $\mathcal{P}_{\rm \Y}$. 
        
        When energy barriers are present, it is harder to take advantage of stochastic fluctuations to carry subsystem \X~forward because it has to be carried over an energy barrier. This produces a peak in output power at intermediate-strength coupling as already described in \cite{Lathouwers2020}. The peak is due to a trade-off as coupling strength increases, between minimizing slip and reducing flexibility for the subsystems to cross the energy barriers one subsystem at a time rather than simultaneously. By contrast, in the barrierless system the heat and power all monotonically change as the coupling strength increases, and notably the maximum output power occurs at tight coupling. Energy barriers consistently slow down the full system, lowering input power, output power, and transduced power compared to the barrierless system.

        How do these trends in transduced power relate to the internal energy and information flows within this multi-subsystem system?  
        One might naively expect that the transduced power between subsystems bounds the output power.
        This would be true if the upstream subsystem’s dynamics were independent of the downstream system, such as when the upstream subsystem is deterministically manipulated by an experimentalist~\cite{Tafoya:2019hr}, or when the coupling energy is negligible compared to the subsystem-specific energetics (weak coupling)~\cite{Large2021}.
        But in autonomous systems where the subsystems are strongly coupled, it turns out to be more complicated because there is capacity transduced through both energetics as well as correlations~\cite{Large2021}: joint entropy also influences the system's capacity to output work.

    \subsection{Bound on input and output powers}
        For bipartite dynamics~\eqref{eq:flux}, the separate subsystem-specific second laws express non-negativity constraints on the subsystem-specific entropy productions.  
        Specifically, applying at steady state the local first laws~\eqref{eq:FLo} to the local second laws~\eqref{eq:EPR} substitutes powers for heats, giving 
        \begin{subequations}
        \label{eq:sub2ndPower}
            \begin{align}
                0 &\le \dot{\Sigma}_{\rm \X} = \beta \mathcal{P}_{\rm \X} - \beta \mathcal{P}_{\rm \X \to \Y} - \dot{I}_{\rm \X} \\
                0 &\le \dot{\Sigma}_{\rm \Y} = \beta \mathcal{P}_{\rm \X \to \Y} + \dot{I}_{\rm \X} + \beta \mathcal{P}_{\rm \Y} \ .
            \end{align}
        \end{subequations}
        
        Rearranging these non-negativity constraints reveals that the output power and input power are each bounded (on opposite sides) by the same quantity:
        \begin{align}
            \beta \mathcal{P}_{\rm \X} \geq \beta \mathcal{P}_{\rm \X \to \Y} + \dot{I}_{\rm \X} \geq -\beta \mathcal{P}_{\rm \Y} \ .
        \end{align}
        This is the same bound identified in \cite{Barato2017} for a bipartite Markov process at steady state.
        
\begin{figure}[!th]
            \centering
            \includegraphics[width=\columnwidth]{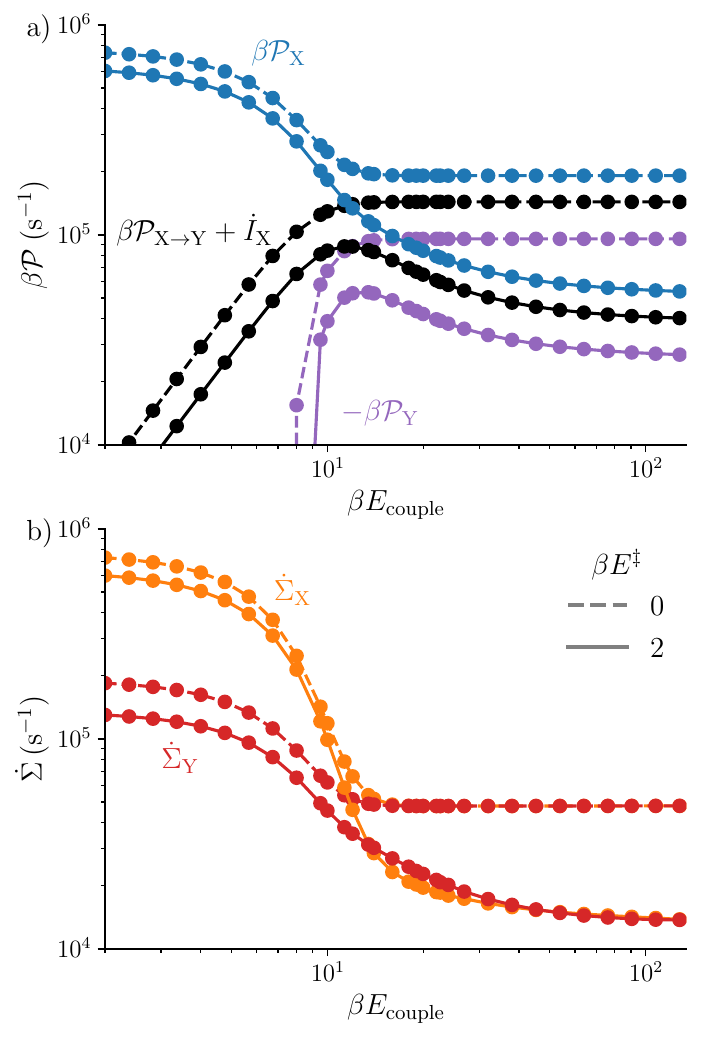}
            \caption{a) 
            Input power $\beta \mathcal{P}_{\rm \X}$, output power $-\beta \mathcal{P}_{\rm \Y}$, and \tc~$\beta \mathcal{P}_{\rm \X \to \Y} + \dot{I}_{\rm \X}$ as functions of coupling strength.
            Input and output power are each bounded by \tc.
            b) Entropy production rates for subsystems \X~and \Y, $\dot{\Sigma}_{\rm \X}$ and $\dot{\Sigma}_{\rm \Y}$ respectively, as functions of coupling strength.
            Dashed curves: barrierless system. Solid curves: system with barriers. 
            Parameters are the same as Fig.~\ref{fig:energyflow}.
            \label{fig:bound}
            } 
        \end{figure}

        This bounding quantity can be no greater than the input power, and places an upper bound on the output power, so we call it the \emph{\tc}.
        The \tc~accounts for the energy and information flows between the subsystems: the transduced power $\mathcal{P}_{\rm \X \to \Y}$ quantifies the change in energy of the combined system due to the dynamics of subsystem \X, and the information flow $\dot{I}_{\rm \X}$ quantifies the change in mutual information between subsystems (which impacts the joint system entropy) due to the dynamics of \X.
        At steady state, \tc~generalizes the transduced additional free energy rate~\cite{Large2021} to the context where the downstream subsystem experiences its own chemical driving force.
        Figure~\ref{fig:bound}a shows the input and output powers, and the \tc~falling in between the two. 
        
        Equations~\eqref{eq:sub2ndPower} also show that the subsystem entropy production rates each equal the difference between the \tc~and the respective power, so as expected they represent the respective losses in capacity at each stage, i.e., due to the dynamics of each subsystem.
        The bound is tightest at tight coupling, and it lies halfway between the input and output powers at the coupling strength that leads to equal EPRs.

    \subsection{Information flow} \label{sub:info}
    
\begin{figure}[!th]
            \centering  
            \includegraphics[width=\columnwidth]{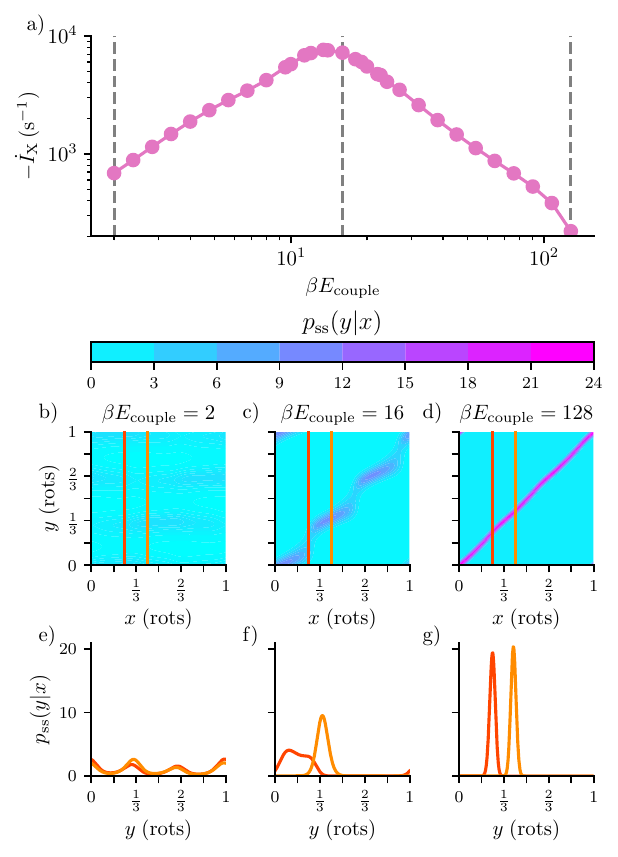}
            \caption{a) Information flow $-\dot{I}_{\rm \X}$~\eqref{eq:infoflow} as a function of coupling strength.
            b)-d) Conditional steady-state probability distribution $p_{\rm ss}(\ey|\ex)$ of the state of \Y~given the state of \X, for b) small ($\beta E_{\rm couple}=2$), c) intermediate (16), and d) large (128) coupling (indicated by vertical dashed lines in a). 
            e)-g) $p_{\rm ss}(\ey|\ex)$ at fixed $\ex = \tfrac{1}{4} \, \rm \add{rotation}$ (red) and $\ex = \tfrac{5}{12} \, \rm \add{rotation}$ (orange), indicated by vertical lines in b)-d).
            Parameters are the same as Fig.~\ref{fig:energyflow}.  
            \label{fig:info_pss}
            } 
        \end{figure}

        The information flow is relatively abstract and generally less familiar than energy flows, so to clarify its meaning in our system, Figure~\ref{fig:info_pss}a) shows the information flow as a function of coupling strength for a system with energy barriers.
        (Barrierless systems have zero information flow, as derived in App.~\ref{app:info}.) 
        The general trend in the information flow is straightforward to understand: at low coupling, each subsystem's motion is independent of the other, leading to a vanishing information flow; by contrast, at high coupling the subsystems' positions are so highly correlated that ongoing dynamics has scarcely any opportunity to provide further information.
        Only intermediate coupling combines dependence of the subsystem states on each other, and new information generation during subsystem dynamics.
        
        The information flow $\dot{I}_{\rm X}$ (Eqs.~\eqref{eq:infoflow2}) can be thought of as the change in uncertainty as \X~evolves about the state of \Y~given the state of \X. 
        If the conditional probability distribution $p(\ey|\ex)$ does not change shape with the conditioning variable \x, then the uncertainty of one variable \y~given the other \x~does not change and the information flow is zero.
        Figures~\ref{fig:info_pss}b-g show that at low coupling (Fig.~\ref{fig:info_pss}b,e), $p(\ey|\ex)$ does not vary with \x; at tight coupling (approximated in Fig.~\ref{fig:info_pss}d,g), $p(\ey|\ex)$ shifts location but does not change shape.
        Both of these extreme cases have unvarying shape of $p(\ey|\ex)$ and hence vanishing information flow.
        Only for intermediate coupling strength (Fig.~\ref{fig:info_pss}c,f), where the shape of $p(\ey|\ex)$ depends on \x, is a non-zero information flow possible.
        In barrierless systems, $p(\ey|\ex)$ is determined only by the coupling part of the system energy (and constant driving forces), making $p(\ey|\ex)$ only a function of difference coordinate $\ex - \ey$, and hence the shape of $p(\ey|\ex)$ does not vary with $\ex$ and the information flow is always zero.
            
        Note that for successful energy transduction, $\dot{I}_{\X}$ is negative and hence $\dot{I}_{\Y}$ is positive: $\X$ must be more responsive to its driving force than to the coupling with $\Y$, hence in general its dynamics reduce information between $\X$ and $\Y$; by contrast, $\Y$ must be more responsive to the coupling than to its driving force, so its dynamics increase information.

    \subsection{Intermediate-coupling features} \label{sub:features}
        Unsurprisingly, the \tc~that upper bounds the output power also peaks at intermediate coupling strength; more surprisingly, the entropy productions are also equal at intermediate coupling strength, near where output power is maximized.  
        Figure~\ref{fig:bound}b shows the subsystem entropy production rates as functions of coupling strength.
        For a barrierless system, the EPR at low coupling is proportional to the square of the driving force applied to each subsystem, and at tight coupling it is proportional to the square of the average force applied to the system, $(\mu_{\rm \X} + \mu_{\rm \Y})/2$.
        With energy barriers, the proportionalities are no longer exact. 
        The barriers suppress the subsystem's motion and lower the EPRs.
        
        In the barrierless case, the EPRs are only equal in the tight-coupling limit, which is also where the output power is greatest. 
        Appendix~\ref{app:EPR} shows analytically that equal EPRs in the barrierless system can be achieved at thermal equilibrium or at tight coupling. 
        For the system with barriers, the EPRs cross at intermediate coupling, before approaching each other again in the tight-coupling limit.
        
\begin{figure}[!th]    
            \centering 
            \includegraphics[width=\columnwidth]{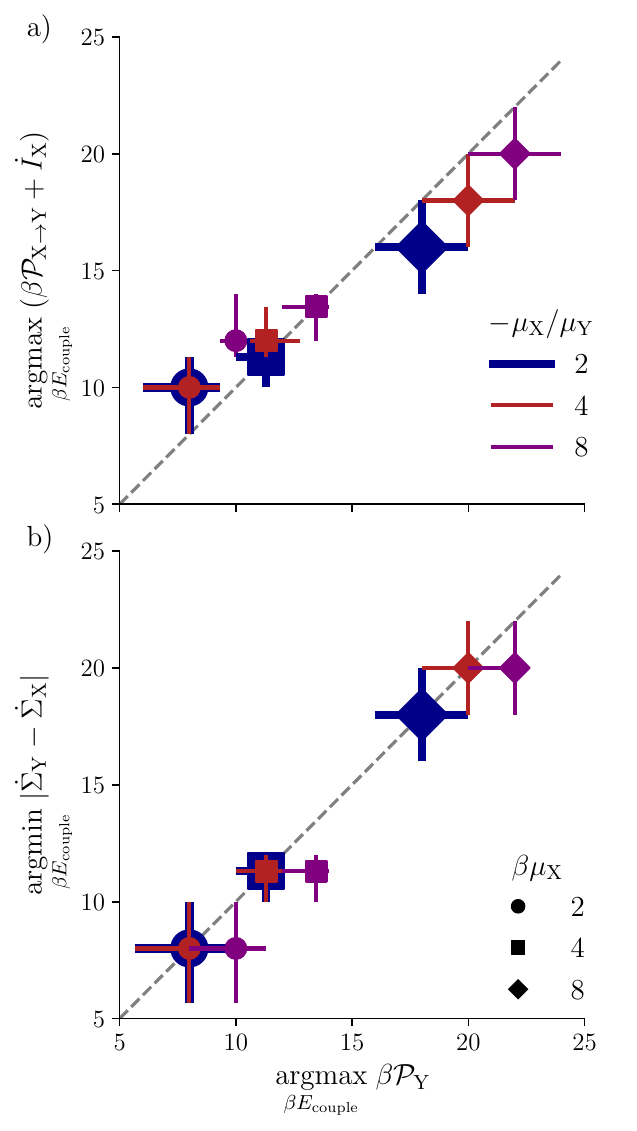}
            \caption{Scatterplots of the coupling strength that maximizes output power and the coupling strength that a) maximizes the \tc~or b) equalizes subsystem EPRs.
            Different colors (blue, purple, red) represent different upstream chemical driving forces, $\beta \mu_{\rm \X} = \{2, 4, 8\} \, 
            \add{\rm rad^{-1}}$. 
            Different symbols (circle, diamond, square) represent different driving-force ratios, $-\mu_{\rm \X}/\mu_{\rm \Y} = \{2, 4, 8\}$.
            Error bars show confidence intervals (between simulated parameter values flanking the maximizing value).
            Blue symbols and error bars are enlarged to ease viewing of overlapping data.}
            \label{fig:correlation}
        \end{figure}

        Figures~\ref{fig:energyflow} and \ref{fig:bound} both show interesting features at intermediate-strength coupling: at around the same coupling strength, the output power $\mathcal{P}_{\rm \Y}$ and \tc~$\beta \mathcal{P}_{\rm \X \to \Y} + \dot{I}_{\rm \X}$ are all peaked and the EPRs are equal.
        To more precisely quantify these observations, Fig.~\ref{fig:correlation} shows the (quite strong) correlation (across different driving forces) between the coupling strength that maximizes the output power and either the coupling strength that maximizes the \tc~or the coupling strength that equalizes the EPRs.
        Appendix~\ref{app:variation} shows the full dependence of these thermodynamic quantities (output power, \tc, and difference of EPRs) on coupling strength.
        
\section{Discussion} \label{sec:discussion}
    In this paper we 
    \add{used a simple model, capturing the essential physics of two strongly coupled stochastic rotary motors, to investigate the effect of coupling strength on internal energy and information flows within a molecular machine and their relation to output power.}
    \stkout{investigated the energy and information flows in strongly coupled rotary motors.}
    In a barrierless system, tight coupling maximizes the output power and transduced power, while the information flow in such a system is zero and the entropy production rates are equal only at tight coupling.
    On the other hand, in a system with energy barriers, intermediate coupling strength maximizes output power $\mathcal{P}_{\rm \Y}$ and \tc\ $\beta \mathcal{P}_{\rm \X \to \Y} + \dot{I}_{\rm \X}$
    \add{(summing the transduced power and information flow from upstream to downstream subsystem)}
    and equalizes subsystem EPRs
    \add{(quantifying subsystem-specific capacity losses)}, 
    roughly the same coupling strength in all cases.
    \add{These findings are rationalized by local subsystem-specific second laws demonstrating that the \tc~forms a lower bound for the input power and an upper bound for the output power.}

    Equal EPRs correspond to dissipation equally distributed over the subsystems.
    Symmetric modeling of the two subsystems---the same energy landscapes with equal numbers of metastable states, equal friction coefficients, equal barrier heights, and coupling that is symmetric around the relative coordinate $\ex - \ey$---likely play a role in the coinciding phenomena such as maximum output power and equal EPRs. 
    In the absence of these symmetries, EPRs need not be equal at any coupling strength, even at tight coupling. 
    It would be interesting to find a condition on subsystem EPRs coinciding with maximal output power in more general models.

    While our model is unique to our knowledge in combining both energy barriers and varying coupling strength in a work-to-work converter, model systems similar to the barrierless special case of our model are quite popular as it can be thought of as diffusion in a tilted periodic potential~\cite{Lee2018,Reimann2001,Lidner2001,Sune2019}, which allows for semi-exact solutions (exact expressions that must be evaluated numerically). Similarly, semi-exact solutions can be found in the tight-coupling limit. 
    For our most general model with arbitrary coupling strength and energy barrier heights, we have only found numerical solution methods.
    
    \add{
    In this work, we focus on systems where each subsystem has $n=3$ metastable states and an identical number of barriers. Preliminary studies suggest that our findings here are qualitatively preserved for smaller and larger $n$~\cite[Sec.~6.4]{LathouwersE_Thesis}, though systematic study of varying $n$ and differing $n$ among the two subsystems would be an interesting extension.
    }
    
    \cite{Fogedby2017} explores a heat engine with a very similar energy landscape.
    \cite{Heinsalu2008,Evstigneev2009} investigate dimer diffusion on a similar landscape, with the main difference being that the particles are subject to the same constant driving force so the system does not function as an engine. 
    \cite{Jack2020} studies thermodynamic uncertainty relations in an identical system to the one presented here, deriving bounds on the EPR and the efficiency and studying the trade-off between dissipation and precision, but they do not comment on the role of coupling strength.
    
    Allosteric interactions, where activity at one location (e.g., ligand binding or conformational rearrangement) affects activity at another spatially separated location, are ubiquitous in molecular motors and many other proteins and protein complexes~\cite{Muneyuki2010,Hilser2012}. 
    Such coupling of distinct functionalities at spatially separated sites is effectively similar to our mechanical coupling, suggesting that our model and its implications may apply to a wide class of energy- and information-transducing systems.

\begin{acknowledgments}
    The authors thank Jannik Ehrich and Matthew Leighton (SFU Physics) for insightful discussions and feedback on the manuscript.
    Special thanks to Joseph Lucero for his integration code for the 
    \stkout{Fokker-Planck}
    \add{Smoluchowski}
    equation.
	This work was supported by a Natural Sciences and Engineering Research Council of Canada (NSERC) Discovery Grant (D.A.S.), a Tier-II Canada Research Chair (D.A.S.), and was enabled in part by support provided by WestGrid (\href{www.westgrid.ca}{www.westgrid.ca}) and Compute Canada Calcul Canada (\href{www.computecanada.ca}{www.computecanada.ca}).
\end{acknowledgments}

\appendix

\section{\add{Computational details}\label{app:compDetails}}
\subsection{\add{Convergence to steady state}}
    \add{Convergence to the steady state is quantified by the total variation distance not changing (to floating-point precision),}  
    \begin{align}
        \frac{1}{2} \int d\ex \int d\ey \, | p(x, y; t + \Delta t) - p(x, y; t) | < 10^{-16} \ ,
    \end{align}
    over a time step of $\Delta t = 10^{-3}$.

\subsection{\add{
Setting the simulation timescale}\label{sec:timescale}}

\add{
    The numerical-simulation timescale is converted to a physical timescale by comparison with the timescale in analogous experiments.
	Crudely approximating ATP synthase as a sphere of radius $r = 15 \, \rm nm$ rotating around an axis through its center, in water with viscosity $\eta = 10^{-9} \ \rm pN  \cdot s \cdot nm^{-2}$, gives a rotational drag coefficient~\cite{Hayashi2010}
	\begin{subequations}
	    \begin{align}
		    \zeta_{\rm r} &= 8 \pi \eta r^3 \\
			&= 2.1 \cdot 10^{-5} \ k_{\rm B} T \cdot \rm s \cdot rad^{-2} \ ,
		\end{align}
	\end{subequations}
	and via the Einstein relation a rotational diffusion coefficient
	\begin{subequations}
	    \begin{align}
		    D_{\rm phys} &= \frac{\kT}{\zeta_{\rm r}} \\
			&= 4.8 \cdot 10^4 \ \rm rad^2 \cdot s^{-1} \ .
		\end{align}
	\end{subequations}
	The simulation diffusion coefficient is set to $D_{\rm sim} = 10^{-3} \ \rm rad^2 \cdot \tau^{-1}$.
	Setting equal the physical and simulation diffusion coefficients, $D_{\rm phys} = D_{\rm sim}$, sets the simulation timescale to
	\begin{align}
	    \tau = 2.1 \cdot 10^{-8}\, \rm s \ .
	\end{align}
}

\section{\add{Derivations}} \label{app:derivations}
\subsection{Transformation of entropy production rates\label{app:entropy}}
    
    In this section, we facilitate \stkout{later SI} calculations \add{in Apps.~\ref{app:info} and \ref{app:EPR}} by deriving the entropy production rates for the transformed coordinates of the center of mass and relative coordinate.
	In the barrierless system with coupling energy that depends only on the relative coordinate $\ex-\ey$, a simple linear transformation decouples the coordinates \x~and \y~into two independent coordinates.
	Consequently, the probability current and the entropy production rate also split into two independent terms.
	
	We start from the definition of the entropy production rate for a bipartite system~\cite{Horowitz2015}, 
	\begin{subequations}
	    \begin{align}
    		\dot{\Sigma} &\equiv \dot{\Sigma}_{\rm \X} + \dot{\Sigma}_{\rm \Y} \\
    		&= \int d\ex \int d\ey \ \frac{\mathcal{J}_{\rm \X}^2(\ex, \ey) + \mathcal{J}_{\rm \Y}^2(\ex, \ey)}{p(\ex, \ey)} \ , 
    	\end{align}
	\end{subequations}
	where the probability current is 
	\begin{align} 
		\mathcal{J}_j(\ex, \ey) = -\frac{1}{\zeta} \left[ p(\ex,\ey) \partial_j V_{\rm eff}(\ex, \ey) + \beta^{-1} \partial_j p(\ex, \ey) \right] \ .
	\end{align}

	The subscript $j$ refers to the probability current for either subsystem \X~or \Y, and similarly $\partial_j$ refers to the related derivative with respect to either \x~or \y. $p(\ex,\ey)$ is the joint probability distribution, and $V_{\rm eff}(\ex,\ey)$ is the effective energy landscape that depends on both subsystem coordinates and chemical driving forces.
	
	We apply a coordinate transformation from \x~and \y~to the center-of-mass \bx~and relative coordinate \dx:
	\begin{subequations} \label{eq:transformation}
	    \begin{align} 
    		\bex &\equiv \tfrac{1}{2}(\ex + \ey), \quad \dex \equiv \ex - \ey \ , \\ 
    		x &= \bex + \tfrac{1}{2}\dex, \quad \ey = \bex - \tfrac{1}{2}\dex \ .
    	\end{align} 
	\end{subequations}
	This transforms the probability distribution,
	\begin{align} \label{eq:prob}
		p(\ex, \ey) = p(\bex) p(\dex) \ ,
	\end{align}
	and the spatial derivatives,
	\begin{subequations}\label{eq:derivatives}
	    \begin{align} 
    		\partial_j p &= (\partial_j \bex) \partial_{\bex} p + (\partial_j \dex) \partial_{\dex} p  \\
    		&= \frac{1}{2}\partial_{\bex} p \pm \partial_{\dex} p \ .
    	\end{align}
	\end{subequations}
	The plus sign refers to the derivative with respect to \x, and the minus sign to the derivative with respect to \y.
	
    The probability currents can then be written in terms of the new coordinates,
	\begin{subequations}
	   \begin{align}
    		\zeta \mathcal{J}_j(\ex, \ey) &= -p(\bex)p(\dex) \, (\tfrac{1}{2}\partial_{\bex} \pm \partial_{\dex} )V_{\rm eff}(\dex) \nonumber \\
    		&\quad - \beta^{-1} (\tfrac{1}{2}\partial_{\bex} \pm \partial_{\dex} ) p(\bex)p(\dex) \\
    		&= \left[ \mp p(\dex) \partial_{\dex} V_{\rm eff}(\dex) \mp \partial_{\dex} p(\dex) \right] p(\bex) \nonumber \\
    		&\quad + \frac{1}{2} \left[ - p(\bex) \partial_{\bex} V_{\rm eff}(\bex) -\partial_{\bex} p(\bex) \right] p(\dex) \\
    		&= \pm \zeta \mathcal{J}_{\rm \dX}(\dex) p(\bex) + \tfrac{1}{2} \zeta \mathcal{J}_{\rm \bX}(\bex) p(\dex) \ , \label{eq:flux2}
    	\end{align}
	\end{subequations}
	where
	\begin{subequations}
	    \begin{align}
    		\mathcal{J}_{\rm \bX}(\bex) &= - p(\bex) \partial_{\bex} V_{\rm eff}(\bex) - \partial_{\bex} p(\bex) \\ 
    		\mathcal{J}_{\rm \dX}(\dex) &= - p(\dex) \partial_{\dex} V_{\rm eff}(\dex) - \partial_{\dex} p(\dex)
    	\end{align}
	\end{subequations}
	are the two independent probability currents.
	\stkout{Thus, we find the sum of squared probability currents is}
    \add{
    Squaring \eqref{eq:flux2} for each of $\mathcal{J}_{\rm X}$ and $\mathcal{J}_{\rm Y}$ expresses the sum of the squared probability currents as
    }
    \begin{align}
		\mathcal{J}_{\rm \X}^2(\ex, \ey) + \mathcal{J}_{\rm \Y}^2(\ex, \ey) =  2\mathcal{J}_{\rm \dX}^2(\dex) p^2(\bex) + \tfrac{1}{2}\mathcal{J}_{\rm \bX}^2(\bex) p^2(\dex) \ .
	\end{align}
	Finally, this is used to calculate the total entropy production rate
	\begin{subequations}
	    \begin{align}
    		\dot{\Sigma} &= \int d\bex \int d\dex \frac{ 2\mathcal{J}_{\rm \dX}^2(\dex) p^2(\bex) +  \tfrac{1}{2}\mathcal{J}_{\rm \bX}^2(\bex) p^2(\dex) }{p(\bex)p(\dex)} \\
    		&= \int d\bex \int d\dex \frac{ 2\mathcal{J}_{\rm \dX}^2(\dex) p(\bex)}{p(\dex)} \\
    		&\quad + \int d\bex \int d\dex \frac{\tfrac{1}{2}\mathcal{J}_{\rm \bX}^2(\bex) p(\dex) }{p(\bex)} \nonumber \\
    		&= 2 \int d\dex \frac{ \mathcal{J}_{\rm \dX}^2(\dex)}{p(\dex)} + \tfrac{1}{2} \int d\bex \frac{\mathcal{J}_{\rm \bX}^2(\bex) }{p(\bex)}  \\
    		&= 2 \dot{\Sigma}_{\rm \dX} + \tfrac{1}{2}\dot{\Sigma}_{\rm \bX} \ .
    	\end{align}
	\end{subequations}

\begin{figure}[th]
        \centering
        \includegraphics[width=0.5\textwidth]{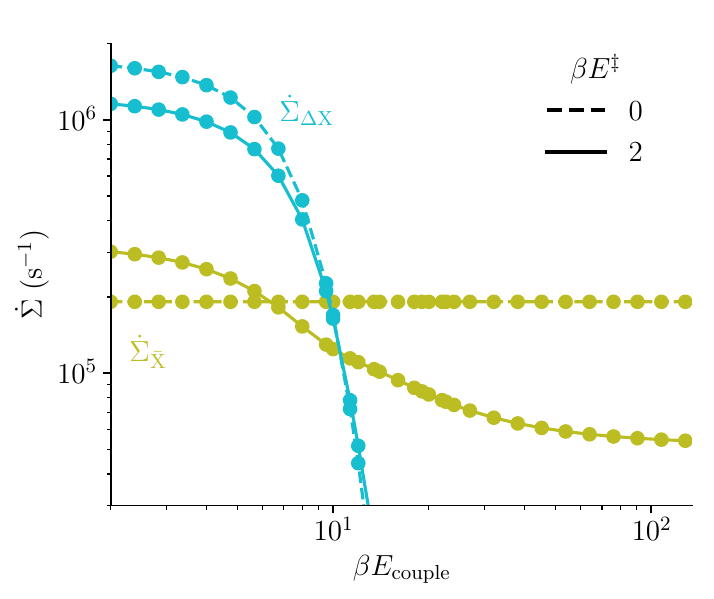}
        \caption{Entropy production rate as a function of coupling strength for the center-of-mass \bx~and difference \dx~coordinates. 
        Dashed curves are for a barrierless system, and solid curves for $2~\kT$ barriers. 
        Parameters are the same as Fig.~\ref{fig:energyflow}.
        \label{fig:EPR}
        } 
    \end{figure}

	Figure~\ref{fig:EPR} shows the entropy production rates for the transformed coordinates \bx~and \dx~as a function of the coupling strength.
	In the tight-coupling limit, the entropy production rate due to the difference coordinate \dx~vanishes, and all of the entropy production is due to the center-of-mass coordinate, regardless of barrier height.
	This is because the probability current through the difference coordinate vanishes as the coupling strength approaches tight coupling.
	This means that in this limit the center-of-mass coordinate is the only coordinate needed to describe the system; it is the reaction coordinate, and the difference coordinate is a bath mode~\cite{Louwerse2021}.
	Finally, note that the EPR in the barrierless case (dashed curves) due to the center-of-mass coordinate \bx~is independent of the coupling strength because the coordinates \bx~and \dx~decouple completely after the coordinate transformation.

\subsection{Calculating information flow} \label{app:info}
    
	Here we show that 
    \add{for our model (with joint energy from Eq.~\eqref{eq:energylandscape}) the information flow is zero (as claimed in Sec.~\ref{sub:info}) when there are no barriers ($V_{\rm X}(x) = V_{\rm Y}(y) = 0$) and the coupling energy}
	\stkout{the information flow is zero (as claimed in Sec.~IV.C of the main text) for our barrierless model with an arbitrary coupling energy that} 
	is only a function of the relative coordinate $\Delta x \equiv \ex - \ey$,
	\begin{align}
		V(\ex, \ey) = V_{\rm couple}(\ex, \ey) = V_{\rm couple}(\ex - \ey) \ .
	\end{align}
	We start from a convenient steady-state expression for the information flow\add{, derived from \eqref{eq:MI} in}~\cite{Allahverdyan2009},
	\begin{align} \label{eq:start}
		\dot{I}_{\rm \X} = -\int d\ex \, d\ey \, \log p(\ey|\ex) \, \partial_{\ex} J_{\rm \X}(\ex,\ey) \ .
	\end{align}
	
	We rewrite the probability current, 
	\begin{align}
		J_{\rm \X}(\ex, \ey) = - \frac{1}{\zeta} \left[p(\ex, \ey) \partial_{\ex} V_{\rm eff}(\ex, \ey) + \beta^{-1} \partial_{\ex} p(\ex, \ey) \right],
	\end{align}
	using the coordinate transformation \eqref{eq:transformation}-\eqref{eq:derivatives} from App.~\ref{app:entropy} to decouple the system dynamics into the probability currents of \bx~and \dx,
    \begin{align}
    	\zeta J_{\rm \X}(\bex,\dex) 
    	&= \zeta \left[ p(\bex) J_{\rm \dX}(\dex) + \tfrac{1}{2} p(\dex) J_{\rm \bX}(\bex) \right].
    \end{align}
	Each of the coordinates' dynamics can be described by a 1D Smoluchowski equation~\cite[Ch.~VIII]{VanKampen1992}. 
	We assume that each of the coordinates, \bx~and \dx, reach a steady state, where the gradient of the probability currents have to equal zero. 
	Consequently, the probability current associated with each of these coordinates has to be uniform\stkout{[S.18]}(independent of the system coordinates),
	\add{
	\begin{subequations}
	    \label{eq:unifFlux}
	    \begin{align}
	    J_{\rm \bX}(\bex) &= C_{\rm \bX} \\
	    J_{\rm \dX}(\dex) &= C_{\rm \dX}.
	    \end{align}
	\end{subequations}
	}
	\stkout{simplifying}
	\add{This simplifies}
	the probability current,
	\begin{align}
	    J_{\rm \add{\X}}(\bex, \dex) &= p(\bex) C_{\rm \dX} + \tfrac{1}{2} p(\dex) C_{\rm \bX} \ .
	\end{align}
	The spatial derivative of the probability current is 
	\begin{subequations}
	    \begin{align}
    		\partial_{\ex} J_{\rm \X}(\bex, \dex) &= \tfrac{1}{2}C_{\rm \dX} \partial_{\ex} p(\dex) + C_{\rm \bX} \partial_{\ex} p(\bex) \\
    		&= \left(\tfrac{1}{2} \partial_{\bex} + \partial_{\dex}\right)\left[\tfrac{1}{2} C_{\rm \bX} p(\dex) + C_{\rm \dX} p(\bex)\right] \\
    		&= \tfrac{1}{2} \left[ C_{\rm \bX} \partial_{\dex} p(\dex) + C_{\rm \dX} \partial_{\bex} p(\bex) \right] \ .
    	\end{align}
	\end{subequations}
	In the second line we changed coordinates of the partial derivative using \eqref{eq:derivatives}.
	Substituting this into \eqref{eq:start}, and transforming the integrals to go over the center-of-mass and relative coordinates, the information flow is
	\begin{subequations}
	    \begin{align}
    		\dot{I}_{\rm \X} &= -\tfrac{1}{2} \int d\ex \, d\ey \, \log p(\ey|\ex) \left[C_{\rm \bX} \partial_{\dex} p(\dex) + C_{\rm \dX} \partial_{\bex} p(\bex) \right] \\
    		&= \tfrac{1}{2} \int d\bex \, d\dex \, \log p(\dex) \left[C_{\rm \bX} \partial_{\dex} p(\dex) + C_{\rm \dX} \partial_{\bex} p(\bex) \right] \\
    		&= \pi C_{\rm \bX} \int d\dex \, \log p(\dex) \, \partial_{\dex} p(\dex) \\
    		&\quad + \tfrac{1}{2} C_{\rm \dX} \int d\dex \log p(\dex) \int d\bex \, \partial_{\bex} p(\bex) \nonumber \\
    		&= \pi C_{\rm \bX} \left\{ p(\dex) [\log p(\dex) - 1] \right\}^{\dex=\pi}_{\dex=-\pi} \\
    		&\quad + \tfrac{1}{2} C_{\rm \dX} \, p(\bex)|^{\bex=2\pi}_{\bex=0} \int_{\Delta x=-\pi}^{\pi} d\dex \log p(\dex) \nonumber \\
    		&= 0 \ .
    	\end{align}
	\end{subequations}
	
	As long as the shape of the conditional probability distribution does not change with the coordinates \x~or \y, the information flow will be zero.
	This result applies to the barrierless system with periodic boundary conditions, in which case the conditional probability distribution is a function of the relative coordinate $\ex - \ey$. 
	\stkout{If, instead, the system does not have periodic boundaries, the arguments presented here need to be adjusted.
	It can be shown that if the relative coordinate \dx~follows a Boltzmann distribution (with no requirement that the center-of-mass coordinate \bx~reaches a steady state or equilibrium), the information flow is also zero.}

\subsection{Equal entropy production rates} \label{app:EPR}

    Here we derive two conditions (as mentioned in Sec.~\ref{sub:features}) in the barrierless system for equality of entropy production rates (EPRs),
	\begin{subequations}
	    \begin{align}
    		\dot{\Sigma}_{\rm \X} &= \int d\ex \int d\ey \frac{\mathcal{J}_{\rm \X}^2(\ex, \ey)}{p(\ex, \ey)} \ , \\
    		\dot{\Sigma}_{\rm \Y} &= \int d\ex \int d\ey \frac{\mathcal{J}_{\rm \Y}^2(\ex, \ey)}{p(\ex, \ey)} \ .
    	\end{align}
	\end{subequations}
	A sufficient, but not necessary, condition for these to be equal is that the square of the probability currents are equal for all \x~and \y. 
	In the barrierless system, the coordinates can be decoupled into two independent coordinates, as in App.~\ref{app:entropy}.
	Additionally, when the coordinates are independent, the steady-state probability current through each coordinate must be uniform\add{~\eqref{eq:unifFlux}.} 
	Using this to rewrite the probability current~\eqref{eq:flux2} where there is nonzero probability gives
	\begin{subequations}
	    \begin{align}
    		\mathcal{J}_{\rm \X}^2(\ex, \ey) &= \mathcal{J}_{\rm \Y}^2(\ex, \ey) \\
    		\left[ p(\bex) C_{\rm \dX} + \tfrac{1}{2} p(\dex) C_{\rm \bX} \right]^2 &= \left[ -p(\bex) C_{\rm \dX} + \tfrac{1}{2} p(\dex) C_{\rm \bX} \right]^2 \\
    		p(\bex) p(\dex) C_{\rm \bX} C_{\rm \dX} &= -p(\bex) p(\dex) C_{\rm \bX} C_{\rm \dX} \\ 
    		C_{\rm \bX} C_{\rm \dX}&= -C_{\rm \bX} C_{\rm \dX} \ . 
    	\end{align}
	\end{subequations}
	This can only be true if both sides equal zero, so one or both of the constants must be zero.
	$C_{\rm \bX} = 0$ implies no probability current through the center-of-mass coordinate; a sufficient condition for this is the center of mass being at thermal equilibrium. 
	Alternatively, $C_{\rm \dX} = 0$ implies no probability current through the relative coordinate; a sufficient condition for this is the tight-coupling limit. 
	Thus tight coupling or thermal equilibrium are sufficient, but not necessary, conditions for equal EPRs.

\section{Variation of central thermodynamic quantities with coupling strength} \label{app:variation}

\begin{figure}[th]    
        \centering 
        \includegraphics[width=\columnwidth]{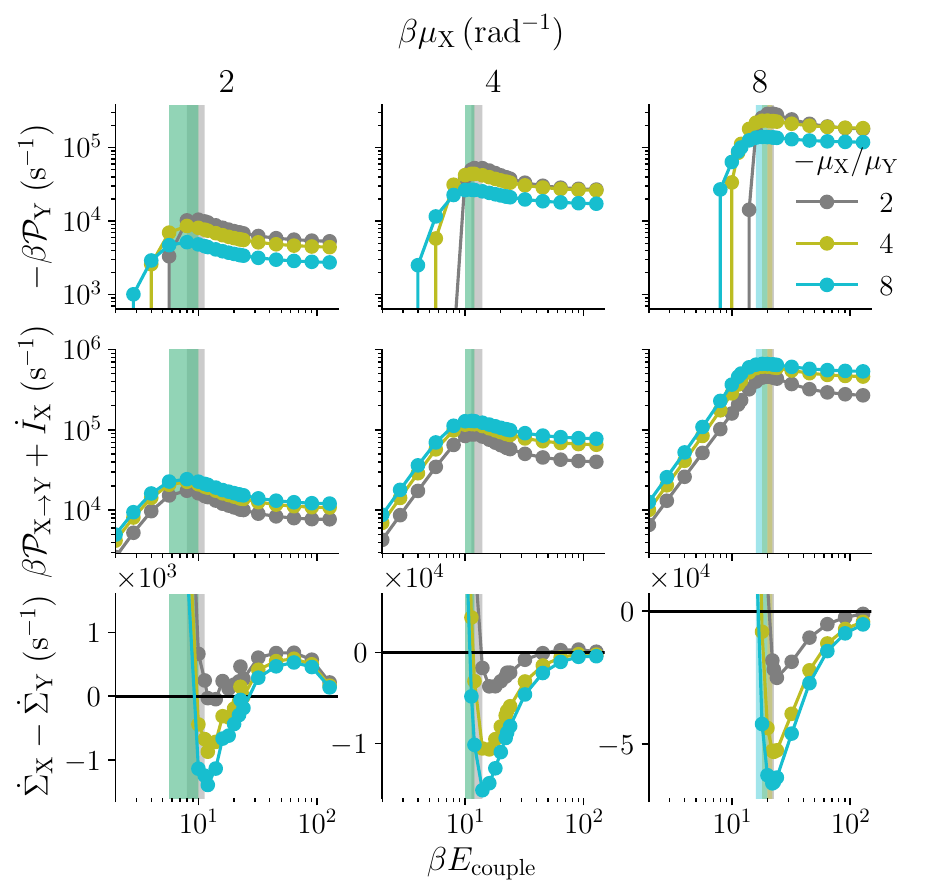}
        \caption{Output power (top row), transduced capacity (middle row), and difference in EPRs (bottom row), each as a function of coupling strength $\beta E_{\rm couple}$, for different driving forces $\beta \mu_{\rm X}$ (left to right columns) and different driving-force ratios $-\mu_{\rm X}/\mu_{\rm Y}$ (different colors).
        Colored bars: confidence intervals for the coupling strength that maximizes the output power.
        \label{fig:grid}
        }
    \end{figure}

    Fig.~\ref{fig:correlation} shows the coupling strength $\beta E_{\rm couple}$ at which the output power $\mathcal{P}_{\rm \Y}$ and the transduced capacity (sum of transduced power $\mathcal{P}_{\rm \X \to \Y}$ and information flow $\dot{I}_{\rm \X}$) are maximized, and at which the subsystem-specific entropy production rates are equal.
    Figure~\ref{fig:grid} shows these quantities' full dependence on coupling strength $\beta E_{\rm couple}$.
    
\eject

\bibliographystyle{apsrev4-1}
\bibliography{references}
	
\end{document}